\documentclass[namedreferences]{solarphysics}
\usepackage[optionalrh]{spr-sola-addons}
\usepackage{epsfig}
\usepackage{graphicx}
\usepackage{color}
\usepackage{url}
\usepackage{lscape}
\usepackage{sidecap}
\newcommand{\etal}{{\it et al.}}
\usepackage[pdfborder={0 0 0 },urlcolor=blue]{hyperref}
\ifx \doiurl \undefined \def \doiurl#1{\href{http://dx.doi.org/#1}{\url{#1}}}\fi
\ifx \adsurl \undefined \def \adsurl#1{\href{http://adsabs.harvard.edu/abs/#1}{\url{#1}}}\fi

\begin{document}
\begin{article}
\begin{opening}

\title{A Study of Variations in Correlation Between Rotation Residual
 and Meridional Velocity of Sunspot Groups}

\author{J.\ Javaraiah}
\institute{Bikasipura, Bengaluru-560 111,  India.\\
Formerly working at Indian Institute of Astrophysics, Bengaluru-560 034, India.\\
email: \url{jajj55@yahoo.co.in;  jdotjavaraiah@gmail.com}; jj@iiap.res.in\\
}
\runningauthor{J. Javaraiah}
\runningtitle{Angular and Meridional Velocities of Sunspot Groups}

\begin{abstract}
We  analyzed  the combined  142 years sunspot-group data
  from Greenwich Photoheliograpic 
Results (GPR) and Debrecen Photoheliographic Data (DPD) and  determined 
the yearly mean residual rotation rate and the meridional velocity of sunspot
 groups in different $5^\circ$ latitude intervals. 
We find that there exists a considerable latitude--time dependence in 
 both the residual rotation and the meridional motion. 
The residual rotation rate  is found to be   $-120$ m $s^{-1}$ to
 $80$ m $s^{-1}$.  In a large number of solar cycles the 
rotation is some extent weaker during maxima than that of during minima.
 There exist alternate bands of equatorward and poleward meridional 
motions. The equatorward motion is dominant mostly around the 
maxima of solar cycles with velocity 8--12 m s$^{-1}$, whereas
the poleward motion is dominant  mostly around the minima but  
 with a relatively weak velocity,  only 4--6 m s$^{-1}$.
The analysis of the data during
Solar Cycles 12--24 that are folded  according to the years from their
respective epochs of maxima suggests
 the existence of  equatorward migrating alternate bands of slower and 
faster than average rotation  within the activity belt.
   This analysis suggests no clear  equatorward or poleward migrating  
  bands of meridional motions.
 A  statistically
significant anticorrelation exists between the meridional motion and residual 
rotation. The corresponding linear-least-squares best-fit is
found to be reasonably good  (slope, $ -0.028 \pm 0.008$, is about 3.5
 times larger than its standard deviation). As per the sign convention used 
for the meridional motions, the  significant negative value of the
 slope indicates the existence of a strong angular momentum transport
 toward equator.  The 
 cross-correlation between the  slopes determined from the data 
in 3-year moving time intervals and yearly mean  
 sunspot number (SN) suggests that the slope leads SN by about 4 
and 9 years.
 The Morlet wavelet spectrum of the slope    suggests  the
existence  of $\approx$11-year periodicity in the slope
 almost throughout the data window, but it was very weak during 1920--1940.
We have also done  cross-wavelet, wavelet-coherence, and wavelet-phase
 difference analyrses of the slope and SN. 
Overall the 
results suggest there exits a strong relationship between the slope 
and amount of activity during a solar cycle. However, 
the correlation  between  the cycle-to-cycle 
modulations in the slope and the amplitude of solar 
cycle is found to be insignificant, indicating that there is no relationship
between the slope and strength of activity on a long-time scale (longer than
 11-year period). 
\end{abstract}


\end{opening}

\section{Introduction}
Studies of variations in solar activity are important for a better
 understanding the basic mechanism of solar activity and solar cycles of 
 various periods to make the long-term forecasts of space-weather and also 
may be  Earth's climate (\opencite{hath15}).
 It is well believed that 
the solar dynamo processes are responsible for solar activity and 
  cycle. The dynamo processes  involve the formation of toroidal
magnetic field due to  the
shearing of the poloidal magnetic field  by differential rotation
and  conversion of the toroidal
 field into poloidal field of opposite polarity  over the
course of approximately 11-year (\opencite{dg06}).
The solar meridional circulation may  play  a vital role 
 to transfer the angular  momentum and magnetic flux across the solar 
latitudes and even can  maintain the observed differential 
rotation (e.g. \opencite{schr85}, and references therein).
Therefore, the studies of the correlation between the
variations in the solar meridional flow
and  differential rotation are important for understanding  role of
 meridional flow in the variations of solar differential rotationi, and hence 
the variations in solar activity.
 The study of correlation between the meridional and rotational
 flows was started by 
\inlinecite{ward65}, who analyzed Greenwich sunspot-group data during the
 period 1935--1944 and found a significant correlation between the angular 
and meridional velocities of sunspot groups and interpreted 
it as meriodional flows transfer angular momentum toward equator.  
Later many scientists
 studied  this correlation by using various data and methods. For example, 
\inlinecite{pat91}  from the Greenwich sunspot data
during the period 1874--1976 and Howard (\citeyear{rf91, rf96})
 using Mt. Wilson 
measurements  of sunspot groups (1917--1985) and of plages (1967--1985)
confirmed the existence of a strong correlation between the meridional 
and rotation velocities of these solar phenomena. 
 However, some authors found the existence of only 
 small covariance of the meridional and angular velocities derived from the 
data of   different tracers (\opencite{nesme93}; \opencite{komm94}; 
\opencite{meun97}).
 Theoretical models (e.g. \opencite{gilm86}) predicted 
the observed correlation between the surface latitudinal and longitudinal 
motions as reflection of equatorial angular momentum transport caused by 
Reynolds stresses near the surface. There were criticisms of  this 
theoretical explanation on the basis that the observed motions may 
result from the well-known expansion and contraction of sunspots along 
the tilted magnetic axes of the sunspot groups. \opencite{gilm84}
argued that because the effect could be observed for whole sunspot groups, 
at least some fraction of the observed correlation must be due to Reynolds 
stresses near the solar surface and that the amount was sufficient to account 
for the angular momentum transport required to maintain the solar 
differential rotation. So far  most of the authors attributed the 
observed correlation between meridional and rotational motions 
 to the action of Reynolds stresses that
demonstrated the presence of a net transport of angular momentum
towards the equator able to maintain the differential
rotation. Recently, \inlinecite{sudar14} by analyzing Greenwich Photographic
 Result (GPR) and Solar Observing Optical Network (SOON) sunspot-group data
 covering the period from 1878 until 2011 and \inlinecite{sudar17}
 by analyzing  the Debrecen Photoheliographic Data (DPD) during the period 
1974--2016, found a statistically significant correlation between 
 sunspot groups'   meridional velocities and   rotation
velocity residuals confirming the transfer of angular momentum towards the 
equator. 

Solar meridional flows vary  during the solar cycle 
 (\opencite{snod87}; \opencite{komm93a}; \opencite{meun05}; 
 \opencite{svanda08}; \opencite{hath10}).
Meridional flows also seem to have a vital role in the cause 
of solar 11-year period torsional oscillations (see \opencite{snod92}) and 
  in solar flux transport dynamo process (\opencite{dg06}). 
\inlinecite{ju06} analyzed a large set of sunspot 
group data (1874--2004) and found  the existence of 
correlation (good in the northern hemisphere and weak in the
southern hemisphere) between the mean solar-cycle variations of meridional
 flow and the latitude gradient term of solar rotation. 
In the present study we analyze  the combined updated  sunspot-group 
data reported in GPR during 
the period 1874--1976 and in DPD during 
the period 1977--2017  and 
study variation in the correlation  between the residual rotation 
and meridional motion of 
sunspot groups and its relationship with solar activity 
 through cross-correlation and continuous- and cross-wavelet 
 analyses. 

In the next section we describe the data analysis, in Section 3
 we show the results, and in  Section~4 we summarize  the conclusions and
 discuss  them briefly.

\section{Data analysis}
 We have downloaded   the  daily sunspot-group data reported
 in GPR during the period
 April 1874\,--\,December 1976
 and  DPD during the period
 January 1977\,--\,June 2017 from the website
{\sf fenyi.\break solarobs.unideb.hu/pub/DPD/}.
The  details about these data can be found in
  \inlinecite{gyr10}, \inlinecite{bara16}, and \inlinecite{gyr17}.
 These
data  contain, beside other
parameters, the date and  time of observation,  heliographic latitude
 ($\lambda$)
and longitude ($L$), and central meridian distance (CMD) of a sunspot group for
each day  during  its appearance on the solar  disk.
 The solar sidereal angular velocity $\omega$ (in degree day$^{-1}$)  and
 meridional velocity  
$v_{mer}$ (in degree day$^{-1}$) of a sunspot group  are calculated 
by using the latitudes and longitudes of the  sunspot group measured at 
times $t_i$ and $t_{i-1}$ during the life time (disk passage) of the 
sunspot group as follows (here $t$ is the date $+$ fraction of the 
day corresponding to the time of observation): 
$\omega(\theta) = \frac{L_i-L_{i -1}}{t_i - t_{i-1}} + 14^\circ.18$ 
and $v_{mer} (\theta) = \frac{\lambda_i-\lambda_{i-1}}{t_i - t_{i-1}}$, 
where $14^\circ.18$ day$^{-1}$ is the Carrington rigid body 
rotation.  
In all our earlier analyses we have
assigned the velocity value to the mean of $\lambda_{i-1}$  and $\lambda_i$.
Following the suggestion by  \inlinecite{ok05}  in \inlinecite{jj20} and 
  here we assigned the velocity value to the $\theta = \lambda_{i-1}$ 
(also see \opencite{sudar14}).
Each disk passage of a
 recurrent sunspot group is  treated as an independent sunspot group.
Hence, we have considered all the sunspot groups that have life times
  2--12 days.  We have not used  the data on the days when 
the $|{\rm CMD}| > 75^\circ$.
 This reduces the foreshortening effect if any. In addition,
 the data  correspond to the absolute  latitudinal
drifts $> 2^\circ$ day$^{-1}$ and absolute longitudinal drifts
 $> 3^\circ$ day$^{-1}$ are excluded.
This reduces considerably the uncertainty in the derived
 results  (\opencite{ward65}; \opencite{jg95}). 
In the case of meridional velocity 
here we have used the following sign convention: 
  in both the northern and  southern hemispheres  positive and negative 
 values indicate the poleward and equatorward motions, respectively. 
We have converted 
the meridional velocity   that measured in degree day$^{-1}$ into  m s$^{-1}$ 
($0.01\ {\rm degree\ day}^{-1} = 1.4\ {\rm  m\ s}^{-1}$, cf.  \opencite{rf91}). 

The data of sunspot groups in the northern and southern hemispheres 
are combined. 
We  binned the daily values of $\omega$ and $v_{mer}$ during each of the 
years 1874--2017  into different $5^\circ$  latitude intervals within 
 $0^\circ$--$40^\circ$ sunspot latitude (absolute) belt.
We determined the  mean values $\langle \omega (\theta) \rangle$
 and $\langle v_{mer} (\theta) \rangle$   of the
daily values of $\omega$ and $v_{mer}$, respectively,  in  each  
$5^\circ$ latitude interval during each year (note that $\langle \ \rangle$ 
implies the mean over a time interval and $\theta$ represents the 
middle value of a latitude interval). 
In several latitudes intervals, particularly in 
a large extent  during the years corresponding to the  
 minima  of the solar cycles, the numbers of velocity values are
found to be zero. In some  
latitude intervals  they are found to be just one. 
The data in all  such latitude 
intervals of all the years are excluded.
We determined mean and standard  error ($\sigma_{mer}$) of 
 $\langle v_{mer} (\theta) \rangle$ of the remaining all latitude intervals
 and in all years.  
The $\langle\omega (\theta) \rangle$ and $\langle v_{mer} (\theta) \rangle$ 
in the  latitude intervals 
 which  correspond to $\langle v_{mer} (\theta) \rangle  > 2.5 \sigma$ are
 excluded. 
We fitted the  values of 
  $\langle\omega (\theta) \rangle$ in the remaining  all latitudes 
intervals during all 
the years 1874--2017  into the standard law of differential rotation:  
$\omega(\theta)  = A + B \sin^2{\theta}$ degree
 day$^{-1}$,  where $A$ and $B$ represent the equatorial rotation rate and 
latitude gradient of rotation, respectively, and
 $\theta$ is the middle value of a 5$^\circ$ latitude interval. 
We obtained the  residual ($\Delta\langle\omega (\theta) \rangle$)  
of the mean  angular velocity $\langle\omega (\theta) \rangle$ in each
 latitude interval (i.e. at each value of $\theta$) by
subtracting the mean value of $\langle\omega (\theta) \rangle$ of 
that latitude interval  from the value 
of $\langle \omega (\theta) \rangle$  of  the same latitude interval  
 deduced from the 
 law of the differential rotation.  We determined the
correlation between   $\langle v_{mer} (\theta) \rangle$ and 
 $\Delta\langle\omega (\theta) \rangle$  
 of all latitude intervals and all years. The corresponding 
   linear-least-squares best-fit is found to be mostly relevant to  the values 
of $\langle v_{mer} (\theta) \rangle  \le 20$ m s$^{-1}$. This is because the 
values of
 $\langle v_{mer} (\theta) \rangle$ that are $>20$ m s$^{-1}$ are 
 few (14\,\%). Hence,    
  we have excluded the  values of $\langle v_{mer} (\theta) \rangle$  and 
$\langle\omega (\theta) \rangle$ in the 
latitude intervals which  are
correspond to the $\langle v_{mer} (\theta) \rangle$$ > 20$ m s$^{-1}$. 
In addition, we have
excluded the values of  $\langle\omega (\theta) \rangle$ and 
 $\langle v_{mer} (\theta) \rangle$  
that  correspond to 
$\Delta \langle \omega (\theta) \rangle > 1^\circ$  day$^{-1}$, 
which are also very
 few (0.6\,\%).
 We repeated all the  aforementioned calculations, 
 i.e. we fitted the remaining   values of  $\langle\omega (\theta) \rangle$  
to the
 standard law of differential rotation and obtained the  differential 
rotation law 
 $\langle \omega (\theta) \rangle  = (14.5 \pm 0.01)-(2.2 \pm 0.07) \sin^2{\theta}$ degree
 day$^{-1}$.  By using the values of $\langle \omega (\theta) \rangle$ deduced 
from this law   we obtained the values of 
  $\Delta\langle\omega (\theta) \rangle$ in all  latitude intervals.  
We have converted  the values of the residual rotation 
$\Delta\langle \omega (\theta) \rangle$ 
degree day$^{-1}$ 
 into  $\Delta\langle v_{rot} (\theta) \rangle$ m s$^{-1}$ to have 
the same units for both  $\Delta\langle v_{rot} (\theta) \rangle$ and
  $\langle v_{mer} (\theta) \rangle$ for 
the sake of convenience to the study correlation between these parameters. 
We determined the 
correlation between $\langle v_{mer} (\theta) \rangle$ and  
$\Delta\langle v_{rot} (\theta) \rangle$ and the  corresponding  
linear-least-squares best-fit
  of these parameters  
in all latitude intervals during the whole period 
1874--2017 and also separately from the values  in each of Solar Cycles 12--24.
We have used the epochs of minima (1878.958, 1890.204, 1902.042, 1913.623, 
1923.623, 1933.707, 1944.124, 1954.288, 1964.791, 1976.206, 1986.707, 1996.624, 2008.958), as well as  maxima (1883.958, 1894.042, 1906.123, 1917.623, 1928.290, 
1937.288, 1947.371, 1958.204, 1968.874, 1979.958, 1989.874, 2001.874, 2014.288)  and the maximum values of
 these cycles  
that were determined by \inlinecite{pesnell18}  by using the 
{13-month smoothed monthly mean} version-2 international sunspot 
number (SN) series.  
We determined  the average behavior of the correlation during 
Solar Cycles 12--24 
by using the method of superposed epoch analysis, i.e. in order to have 
a better statistics  all the 13 solar cycles 
data are superposed/combined  according to the years from the respective 
epochs of maxima of the solar cycles. Such analyses were done by many 
authors~(\opencite{gilm84}; \opencite{balth86}; \opencite{jk99};
\opencite{jj03}; \opencite{ju06}; \opencite{braj06}; \opencite{sudar14}). 
We have also  binned the  values of $\Delta\langle v_{rot} (\theta) \rangle$ 
and   $\langle v_{mer} (\theta) \rangle$  into 3-year moving time 
intervals (3-year MTIs) successively 
shifted by one year, namely 1874--1876, 1875--1877,\dots,2015--2017 and  
determined the corresponding correlation and the linear-least-squares
  best-fit   in each of the 3-year MTIs. 

Wavelet transform can be used to analyze time
series that contain nonstationary power at many different
frequencies. That is, by wavelet analysis it is possible to 
 decompose a time series into time--frequency
space. Hence, one can determine both the dominant
modes of variability and how those modes vary
in time. The cross-wavelet transform between two time series is
simply the product of the first complex wavelet 
transform with the complex conjugate of the second.
The cross-wavelet power spectrum can be used as a quantified 
indication of the similarity of power between two time series.
The wavelet-coherency  is 
a normalized time and scale (period) resolved measure for the
relationship between two time series. It is 
  the square of the cross-wavelet power spectrum
normalized by the individual wavelet-power spectra.
This gives a quantity between 0 and 1. 
A value of 1 means the existence of linear relationship between the 
two time series  around a time on a scale. A value of zero 
means vanishing correlation (no linear relationship). 
Measurements of wavelet-phase difference between two time series
yield information on the phase delay between oscillations in
the time series as a function of frequency (for details see \opencite{tc98}).
We have done  Morlet wavelet analysis for the time series of 
the slopes of the linear relationships 
 in 3-year MTIs. We have also done the wavelet analysis for  the yearly 
values of the version-2 SN.
The yearly average  version-2  SN  time series is  
  downloaded from  {\sf www.sidc.be/silso/datafiles}.
The details of changes and corrections in version-2 SN can be found in
\inlinecite{clette16}. 
Similarities in the Morlet wavelet spectra of the slope and SN are 
checked from cross-wavelet,  wavelet-coherence, 
 and wavelet-phase difference analyses. 
We have used the IDL-codes of the wavelet 
analyzes provided by  \inlinecite{tc98}  and available at 
{\tt paos.colorado.edu/research/wavelets}.

\begin{figure}
\centerline{\includegraphics[width=\textwidth]{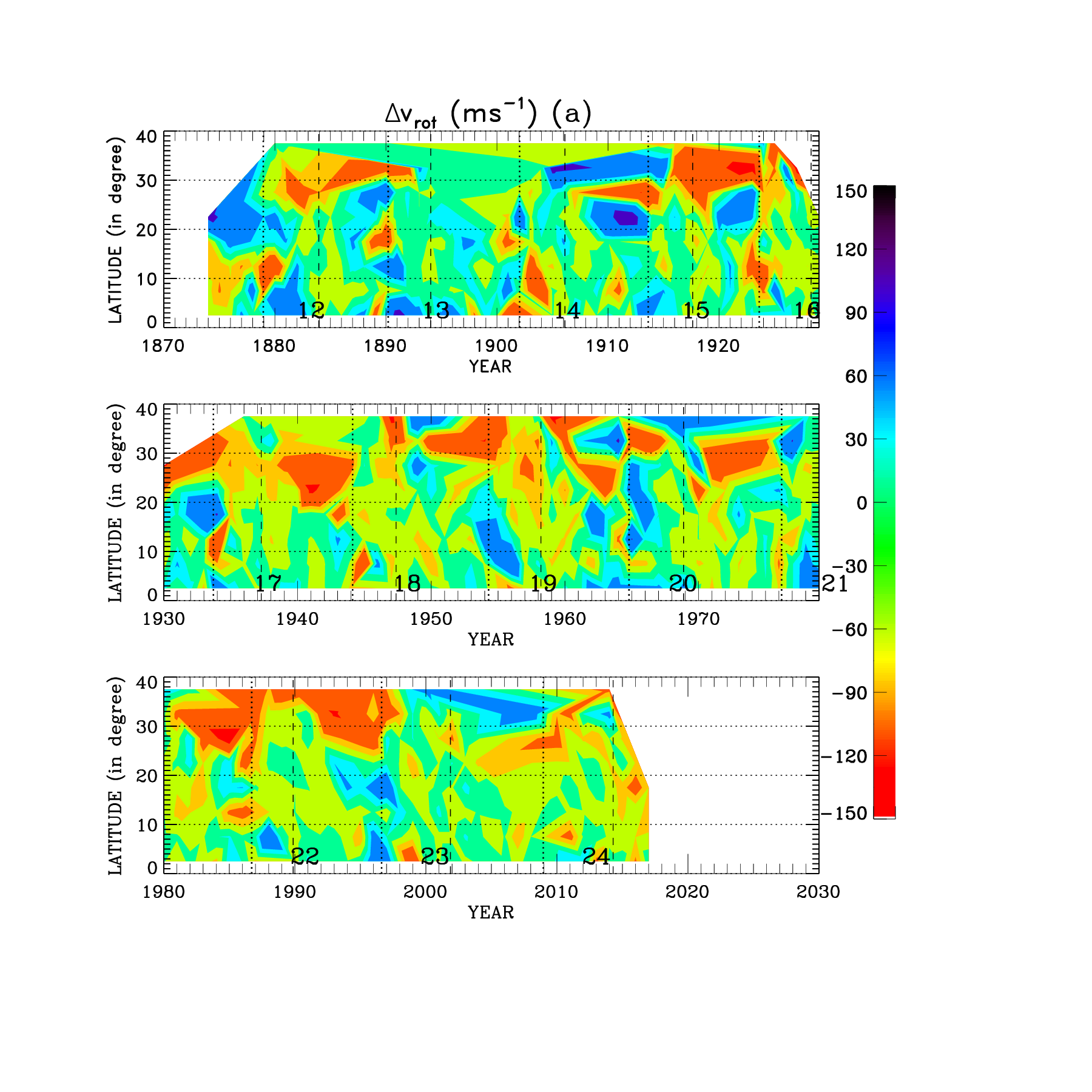}}
\vspace{-2.5cm}
\centerline{\includegraphics[width=\textwidth]{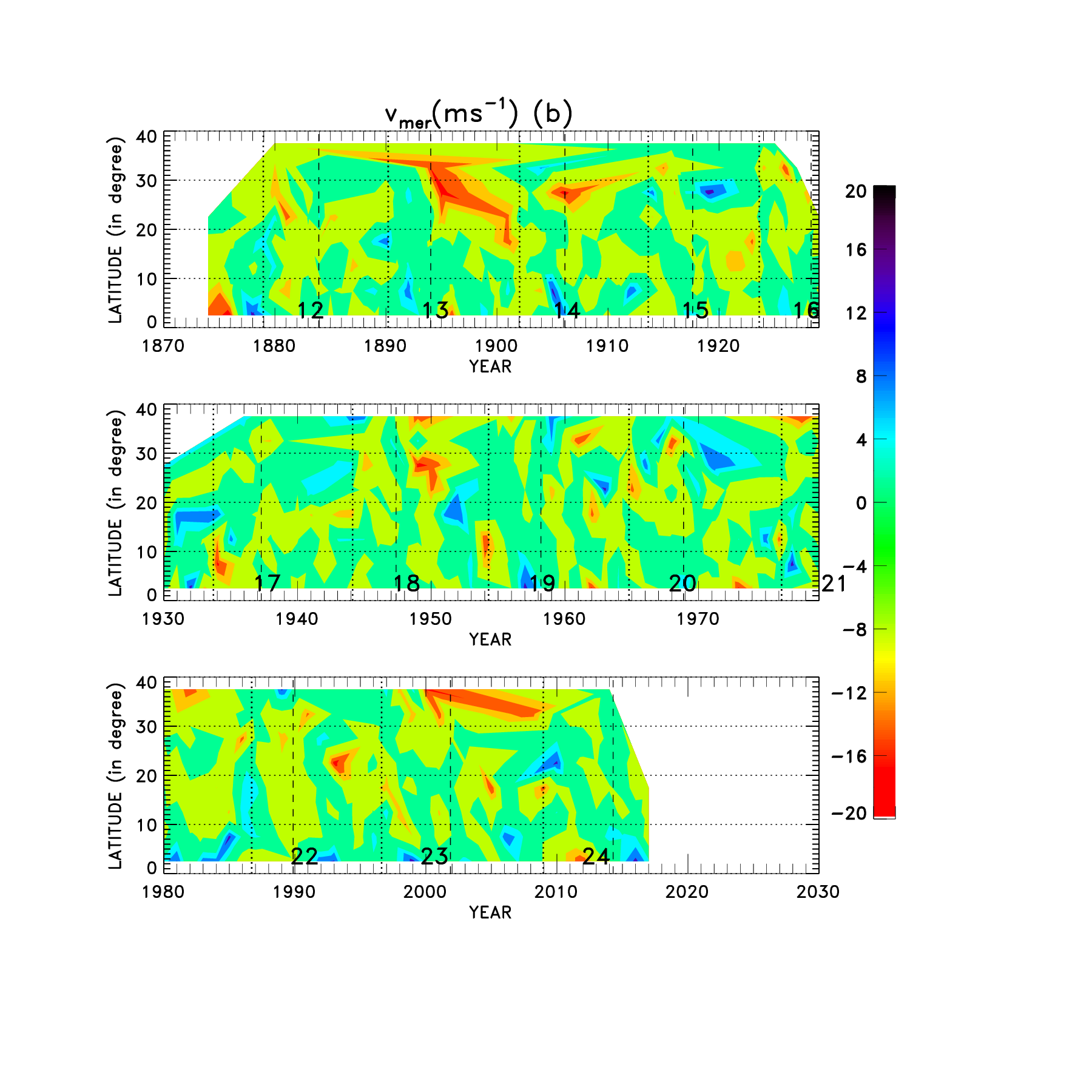}}
\vspace{-2.0cm}
\caption{Variations in  ({\bf a}) yearly mean
  residual rotation rate ($\Delta\langle v_{rot} (\theta) \rangle$)
and  ({\bf b}) yearly mean   meridional 
velocity ($\langle v_{mer} (\theta) \rangle$) 
in different $5^\circ$ latitude intervals ($\theta$ is the middle value
of a latitude interval) during 1874--2017. 
Northern and southern hemispheres' data are folded. 
Positive and 
negative values of meridional velocity represent poleward and  
equatorward motions, respectively.} 
\label{f1}
\end{figure}

\begin{figure}
\centerline{\includegraphics[width=\textwidth]{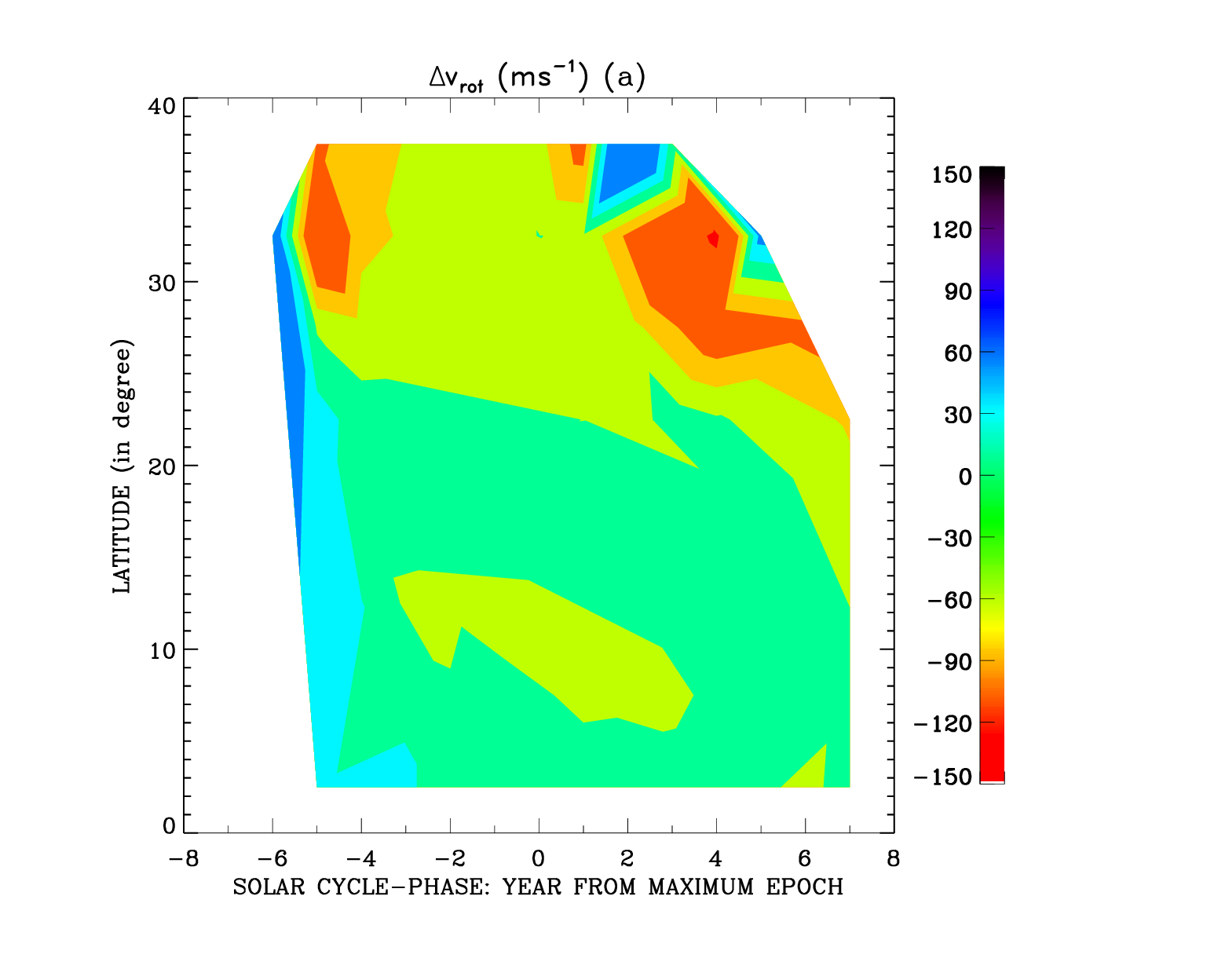}}
\vspace{-1.0cm}
\centerline{\includegraphics[width=\textwidth]{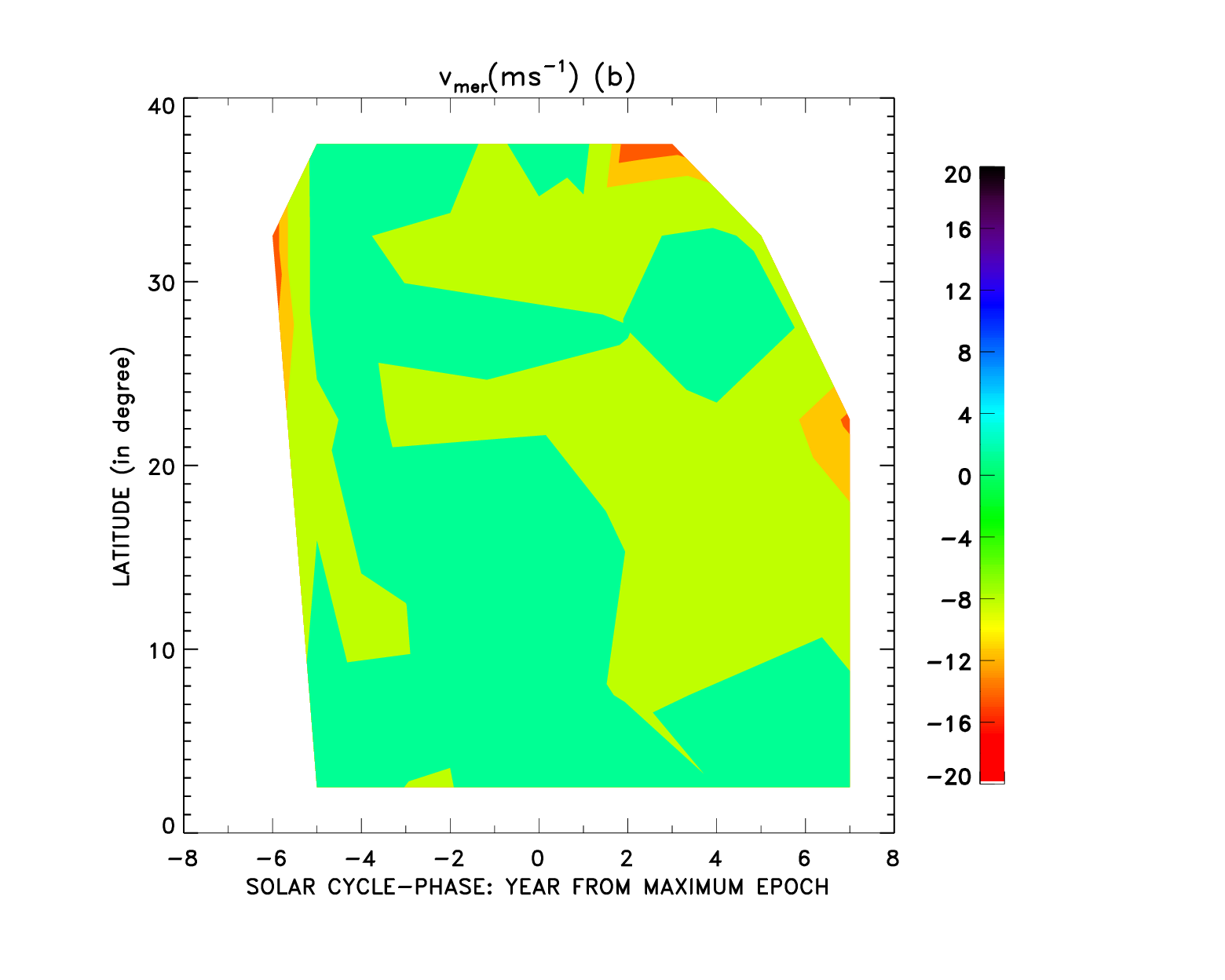}}
\vspace{-0.8cm}
\caption{Average variation in  ({\bf a}) yearly mean  residual
 rotation rate ($\Delta\langle v_{rot} (\theta) \rangle$)  
and ({\bf b})  in yearly mean meridional velocity 
($\langle v_{mer} (\theta) \rangle$)  of sunspot groups 
in different $5^\circ$ latitude intervals  during  
different phase of Solar Cycles 12--24, 
determined by superposing  the  yearly data 
(the values shown in Figure~1)  of the 
Solar Cycles 12--24 (Cycle 24 is incomplete)  
according to the years from their 
respective epochs of maxima (the epochs of maxima are taken as zeros). 
Northern and southern hemispheres' data are folded. Positive and 
negative values of meridional velocity represent poleward and  
equatorward motions, respectively.} 
\label{f2}
\end{figure}

\section{Results}
\subsection{Latitude-Time Dependence of $\Delta\langle v_{rot} \rangle$ and 
$\langle v_{mer} \rangle$}
Figure~1a shows the variations in  yearly mean residual rotation 
($\Delta\langle v_{rot} (\theta) \rangle$)  whose values  correspond to the 
$\Delta \langle \omega (\theta) \rangle  \le 1^\circ$ day$^{-1}$ 
($\approx$ 140 m s$^{-1}$), 
and Figure~1b shows yearly mean meridional 
velocity ($\langle v_{mer} (\theta) \rangle$)
 which has values only
 $\pm 20$ m s$^{-1}$, of sunspot groups
in different $5^\circ$ latitude ($\theta$) intervals during 1874--2017.

As can be seen in Figure~1a,  there 
is a considerable latitude--time dependence in 
$\Delta\langle v_{rot} \rangle$.
The values of  $\Delta\langle v_{rot} (\theta) \rangle$ are
from  $-120$ m $s^{-1}$ to
 $80$ m $s^{-1}$ (only a few values are beyond this interval).
 In a large number of solar cycles the 
rotation is some extent weaker during maxima than that of during minima.  
There is an  indication 
on high-to-low latitude  migration  in  
  $\Delta\langle v_{rot} (\theta) \rangle$ over 8--10-year period. 
Particularly, the band of slower than average rotation seems to be 
migrating  from $\approx35^\circ$ latitude to around $5^\circ$ 
latitude  during many solar cycles  (not clearly visible  
in the early solar cycles).

As can be seen in Figure~1b, there is also a 
considerable latitude--time dependence in $\langle v_{mer} \rangle$. 
There seem to be  alternate bands of equatorial motion
 ($\langle v_{mer} (\theta) \rangle$ is negative)
 and poleward  
motion ($\langle v_{mer} (\theta) \rangle$ is positive).
 Overall, the equatorward 
motion is
 dominant with 
velocity 8--12 m s$^{-1}$ mostly around maxima of more solar cycles, 
whereas the poleward motion seems to be relatively weak 
with velocity  only 4--6 m s$^{-1}$ and exists mostly around  minima of 
more solar cycles (also see \opencite{ju06}). 
 However, in some solar cycles, e.g. 18 and 24, the motion 
was seem to be equatorward during their whole  periods.
There is also some indication of 
 equatorward  migration   in the latitude--time
 dependence of $\langle v_{mer}\rangle$ during a large number 
of solar cycles. The beginnings and endings of the 
equatorward bands are not at exactly the beginnings and endings of 
solar cycles.  
Poleward meridional flow with a speed of about 20 m s$^{-1}$ 
 has been well established at the
 solar photospheric level (\opencite{meun09}).

Figures~2a and 2b show the  
average variations in yearly mean $\Delta\langle v_{rot} \rangle$
and $\langle v_{mer} \rangle$  of sunspot groups
in different $5^\circ$ latitude intervals  during 
 Solar Cycles 12--24. This is 
determined by superposing the  yearly data (the values shown in Figure~1)  of
Solar Cycles 12--24  according to the years from their
respective epochs of maxima. 
Northern and southern hemispheres' data are folded. 
Figure~2a shows the existence of alternate bands of slower and faster 
than average rotation, i.e. alternate bands of  negative and positive 
values of  $\Delta\langle v_{rot} (\theta) \rangle$,  within the activity 
belt. 
 The about $10^\circ$  wide slow band (residual rotational 
velocity is  $-$70 to $-80$ ms$^{-1}$)  seems to be
 originated around $35^\circ$ latitude and  the narrow one
(only about $5^\circ$ wide)  seems to be 
 originated around  $15^\circ$ latitude during the minimum of
 a solar cycle and both migrated toward low latitudes. The latter looks 
to be ended 
in the declining  phase (before end) of  solar cycle. The speed of
 migration of former is 
much low up to middle of the declining phases of solar cycles
 and then suddenly increased. It should be noted here that there is 
a major drawback in  a superposed epoch analysis. The lengths  
 and rise times of solar cycles are considerably different. 
The  values at beyond the epochs $-4$ and $4$ (cf. Figure~2a)   
 represent for a few solar cycles rather than an average of 
all solar cycles. In a similar analysis \inlinecite{sudar14} 
did not see any regularity in changes of  pattern in
 $\Delta\langle v_{rot} (\theta) \rangle$ over time.     
 
As can be seen in Figure~2b  there are bands of equatorward meridional 
motion ($\langle v_{mer} (\theta) \rangle$ is negative), separated 
by a band of poleward motion in $26^\circ$--$29^\circ$ latitude interval.
These bands  
equatorward or poleward migrations are not clearly visible.
 The motions of active regions largely represent the
plasma motion in the sun's subsurface layers (\opencite{rf96}). 
 During the whole  decline phase and in the center of activity 
belt the motion is mostly equatorward.
Around the maxima of solar cycles at the middle and low latitudes 
the motion is mostly poleward and in $21^\circ$--$25^\circ$  and 
$30^\circ$--$35^\circ$ latitudes the motion is 
 equatorward. Close to the beginnings (epoch $-4$) of solar cycles,  
the motion looks  to be mostly poleward at all latitudes.  
Since the magnetic structures of large/long-lived sunspot groups 
might anchor relatively deeper than those of small sunspot groups 
(see \opencite{jg97}), the equatorward motions of sunspot 
groups during  maxima of solar cycles represent the 
 motions in  relatively deeper layers of the sun than 
 the poleward motions of sunspot groups during  minima of solar cycles and 
mostly in low latitudes. 
 The overall pattern of the motion is somewhat consistent with the concept of 
flux-transport dynamo models (\opencite{dg06}).  
\opencite{siva10} analyzed the Mt. Wilson and Kodaikanal
 sunspot-group data and  found the equatorward motion in all latitudes.
These authors have used only first days' data of sunspot groups, whereas 
here we have used the data in all days during the life times of the sunspot 
groups. \inlinecite{zhao04} by employing a time–distance technique of 
helioseismology found that the residual meridional flows (the flows
 subtracted by the mean meridional flow profile of 1996) converged 
toward the solar activity belts. Recently,  \inlinecite{sudar14} noted 
that  the meridional motion of sunspot groups is directed towards
 the zone of solar activity. This result  is largely confirmed 
 here around the maxima of solar cycles (see Figure~2b),  however, 
 the poleword motions (4--5 ms$^{-1}$)  are much weaker than the  
equatorword motions (8--10 ms$^{-1}$).
\inlinecite{rf91} found that sunspots groups tend to move away 
from the central latitude of activity. This is  not found  here.

\subsection{Comparison of Variations in  
 $\Delta\langle v_{rot} (\theta) \rangle$ with Torsional Oscillations}
The so-called torsional oscillation discovered by \inlinecite{hl80}
from  Mt. Wilson velocity data during the period 1967--1980 
consists of alternating bands of faster (or 
slower) than average rotation moving from high latitudes towards 
the equator in $\approx$22-year time. The maximum amplitude of the torsional 
oscillation is about 5 m s$^{-1}$.
The faster-than-average 
rotation band is located on the equatorward side of the magnetic 
activity belt and a slower-than-average rotation band is located 
on its poleward side. Here, obviously, both the 
 slower and faster than average rotation, i.e. bands of 
negative and positive values of $\Delta\langle v_{rot} (\theta) \rangle$, 
exist within the activity belt. 
The torsional oscillation pattern was also found from full-disk 
magnetograms but with the maximum of amplitude
of the magnetic pattern is twice as large as that of the
Doppler pattern (\opencite{snod91}; \opencite{komm93b}). 
The maximum absolute value 
of $\Delta\langle v_{rot} (\theta) \rangle$, that is determined here from
sunspot-group data,  is much larger than the amplitude of the  velocity
 as well as the  magnetic  torsional oscillations. It is  
well-known that magnetic structures of sunspots are rotating  2--3\,\%  
faster than the surrounding plasma (see \opencite{jk02}). This is commonly  
interpreted as the 
magnetic structures anchored at different depths
in the convective zone, being coupled to layers rotating at a different
 velocity (see \opencite{jg97}). 
We know from helioseismology that there is a rotation  gradient just below
 the surface (\opencite{antia02}). 

It has been claimed that the torsional  pattern is present in 
sunspot rotation  (\opencite{gm82}; \opencite{ttk83}).
 \inlinecite{ttk83}  analyzed longitudinal and latitudinal motions of 
recurrent sunspot groups using Greenwich 
 data during 1874--1976.  Although their time resolution was coarse, 
they found some evidence for the 11-yr oscillation in the sunspot 
zones with an amplitude of a few m s$^{-1}$. \inlinecite{gilm84} 
 and \inlinecite{balth86} 
observed faster and slower bands, but with no clear migratory 
character.  \inlinecite{tern90}  found an evidence of equatorward 
moving bands of torsional oscillation through a very careful study of 
the sunspot drawings 
made during Cycle 21 at Catania Astrophysical Observatory. 
The latitude-time dependent pattern of  
$\Delta\langle v_{rot} (\theta) \rangle$ that is seen in Figures~1a and 2a
is largely similar to that aforementioned pattern. 
 \inlinecite{tern90} used 
only the data of old sunspot groups, i.e. for each sunspot group 
only the data collected from the 4th day of observation until the last 
observation available have been taken into account. Here we have used 
all the available data  of sunspot groups whose life times were 2--12 days.

 Meunier et al. (1997)  analyzed the rotation of 
photospheric faculae obtained 
at Meudon throughout Cycle~19 (1954--1964) and found  
bands of faster and slower 
rotation rates with an amplitude of a few meters per second similar 
to the  torsional oscillations. 
 \inlinecite{maka97} studied 
the long-term variations of the differential rotation of the solar 
large-scale magnetic field using synoptic H$\alpha$ maps in the 
latitude zone from $+ 45^{\rm o}$ to $-45^{\rm o}$ in the period 
of 1915--1990. In each solar cycle, they found a band of faster or slower 
than average rotation moving from high to low latitudes. The slow 
band roughly corresponds to the location of magnetic activity, which is 
 in a large extent similar  as  the patterns in  
$\Delta\langle v_{rot} (\theta) \rangle$ are seen here. 

\begin{SCfigure}
\centering
\includegraphics[width=7.5cm]{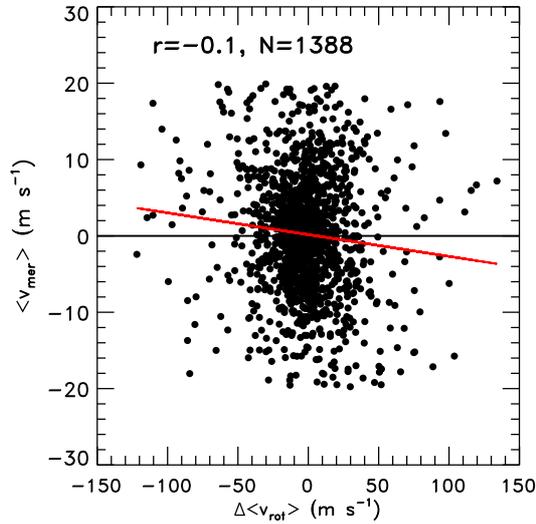}
\caption{Scatter plot of meridional velocity 
($\langle v_{mer} (\theta) \rangle$ from $-20$ to $+20$ m s$^{-1}$) 
 versus residual rotation rate
 ($\Delta \langle v_{rot} (\theta) \rangle$  $\approx -140$ 
 to $+140$ m s$^{-1}$) 
 determined by the combined data of sunspot groups during 
Solar Cycles~12--24  ($r =-0.1$, number of data points $N = 1388$, 
 Student's $\tau = 3.73$, $P = 0.001$). 
Positive values  of meridional velocity indicate poleward motions and negative
 values indicate equatorward motions. 
The  continuous line (red) represents the linear best-fit  
($ {\rm slope} =  -0.028 \pm 0.008$)}.
\label{f3}
\end{SCfigure}

\subsection{Correlation Between $\langle v_{mer} (\theta) \rangle$  and
  $\Delta\langle v_{rot} (\theta) \rangle$} 
Figure~3 shows the correlation between
 $\langle v_{mer} (\theta) \rangle$  and
  $\Delta\langle v_{rot} (\theta) \rangle$ 
 determined from the whole data gone in Figures 1a and 1b 
(the combined data of sunspot groups during
Solar Cycles~12--24). As can be seen in this figure  a statistically 
significant anticorrelation  between $\langle v_{mer} (\theta) \rangle$  and  
$\Delta\langle v_{rot} (\theta) \rangle$ exists 
(also see \opencite{sudar14}). 
The correlation coefficient ($r$) is $-0.1$, which is significant at
the 99.9\% 
confidence level  ($N = 1388$, Students' t: $\tau = 3.73$, $P = 0.001$).
The corresponding linear-least-squares fit is reasonably good in the sense 
that the slope  ($ -0.028 \pm 0.008$) is about 3.5 times larger than its
 standard 
deviation. Based on the sign convention used for meridional 
velocities of sunspot groups (Sect. 2 above), the negative value of 
the slope can be considered as a measure of 
the strength of angular momentum transport toward equator.

\begin{table}
{\scriptsize
\caption[]{The values of intercept ($C$)  and slope ($D$) of the linear
relationship between $\langle v_{mer} (\theta) \rangle$ and  
$\Delta\langle v_{rot} (\theta) \rangle$ at each epoch (year) 
determined from the  yearly mean  values  of 
Solar Cycles 12--24  superposed according to the years from their 
respective epochs of maxima, i.e. determined from the values shown Figure~2. 
The corresponding values of the correlation coefficient ($r$), 
Students' t ($\tau$), probability ($Prob.$), and number of data points ($N$)
 are also given. The corresponding  values of all these
 parameters   determined from the combined values of 
$\langle v_{mer} (\theta) \rangle$ and  
$\Delta\langle v_{rot} (\theta) \rangle$ of all the latitude 
intervals  during the entire period, $-6$ to $8$, are given in the last row.}
\begin{tabular}{lccccccccccccccccccccccccccccccccc}
\hline
Epoch & $C$ & $D$& $r$ & $\tau$& $Prob.$&  $N$\\
 $-6$&$  -0.67\pm   5.13$&$  -0.32\pm   0.18$&$  -0.87$&   1.80&  0.838&    5\\
 $-5$&$   0.74\pm   1.69$&$   0.06\pm   0.04$&$   0.26$&   1.30&  0.896&   27\\
 $-4$&$   1.93\pm   1.04$&$  -0.06\pm   0.03$&$  -0.26$&   2.13&  0.982&   69\\
 $-3$&$  -0.83\pm   0.77$&$   0.00\pm   0.02$&$   0.02$&   0.18&  0.573&  124\\
 $-2$&$   1.33\pm   0.70$&$  -0.01\pm   0.02$&$  -0.03$&   0.37&  0.643&  148\\
 $-1$&$   0.19\pm   0.56$&$  -0.04\pm   0.02$&$  -0.15$&   1.90&  0.970&  169\\
  0&$  -0.09\pm   0.52$&$  -0.07\pm   0.02$&$  -0.22$&   2.98&  0.998&  172\\
  1&$   0.72\pm   0.54$&$   0.00\pm   0.03$&$  -0.01$&   0.07&  0.530&  161\\
  2&$  -0.75\pm   0.58$&$  -0.13\pm   0.03$&$  -0.39$&   5.05&  1.000&  149\\
  3&$  -0.76\pm   0.63$&$  -0.05\pm   0.03$&$  -0.13$&   1.46&  0.927&  127\\
  4&$  -1.11\pm   0.73$&$  -0.08\pm   0.03$&$  -0.26$&   2.86&  0.997&  113\\
  5&$   0.32\pm   0.83$&$  -0.04\pm   0.04$&$  -0.12$&   1.06&  0.855&   84\\
  6&$   0.64\pm   1.01$&$   0.05\pm   0.03$&$   0.23$&   1.95&  0.972&   70\\
  7&$   1.83\pm   2.39$&$   0.15\pm   0.07$&$   0.44$&   1.98&  0.967&   20\\
Whole&$-0.27\pm   0.74$&$  -0.085\pm   0.028$&$  -0.40$&   3.06&  0.998&   50\\
\hline
\end{tabular}
\label{table1}
}
\end{table}

We also determined  correlation
between $\langle v_{mer} (\theta) \rangle$  and
$\Delta\langle v_{rot} (\theta) \rangle$ at each epoch (year) in 
Figures~2a and 2b, i.e. during the average solar cycle of the 
superposed data of Solar Cycles
12--24.  We have done  linear-least-squares fit to the  
corresponding data.   In Table~1 we have given the values of the 
corresponding parameters, 
 namely correlation coefficient ($r$), the slope of best-fit 
 linear relationship, etc. 
In many years the value of $r$ found to be significant at above 95\% 
confidence level ($Prob. \ge 0.95$, i.e., $P \le 0.05$), obtained from 
Student's t-test.  The correlation of  
  the combined data of all epochs is also found to be statistically 
significant (see the last row in Table~1).
 Figure~4 shows the 
variation in the slope during the average solar cycle. 
 If we exclude the point at the epoch $-6$, a 11-year period 
 cycle pattern 
(anticorrelation with solar cycle) seems to present in the slope, 
suggesting that there  exists a relationship between the slope and
activity during solar cycles (also see \opencite{sudar14}). 
 However, the  positive value  of the slope that corresponds
 to the minimum epochs of 
a few long cycles have a large uncertainty 
($\sigma$, standard deviation)

\begin{SCfigure}
\centering
\includegraphics[width=7.5cm]{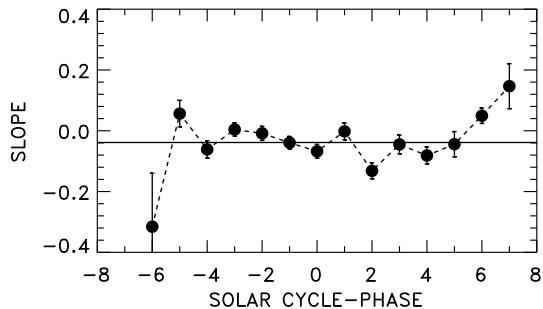}
\caption{Mean variation of the  slope 
of the linear best-fit of meridional velocity 
($\langle v_{mer} (\theta) \rangle$) and residual rotation  
 ($\Delta\langle v_{rot} (\theta) \rangle$)  during Solar Cycles 12--24 
determined from the superposed epoch analysis (used the values of the 
slope given Table~1). Zero represents the maximum epoch of a solar
 cycle.}
\label{f4}
\end{SCfigure}

Figure~5a shows the variation in the slopes of the linear relationships 
of $\langle v_{mer} (\theta) \rangle$  and 
$\Delta\langle v_{rot} (\theta) \rangle$ determined 
from the data in 3-year MTIs
during the period 1874--2017.  As in some of our earlier analyses 
(e.g. \opencite{jk99}) we have revised the time series of the slopes 
in 3-year MTIs by replacing those values of the slopes  having 
 $\sigma$ greater than
 1.7 times the median-$\sigma$  with the mean of their respective
neighbor values.  They are at 1878, 1896, 1907, 1908, 1932, 1952, 1974,       2005, 2006, 2008, 2009, and 2010. (No value of $\sigma$ is found to 
be greater than 2.5 times the median-$\sigma$.)  In Figure~5a we have also 
shown the variation in 
 yearly mean values of SN  during 1875--2016 
(used the file SN\_y\_tot\_v2.0.txt that was downloaded from
 {\sf www.sidc.be/silso/datafiles}). 
In Figure~5b we have shown the cross-correlation between the 
slope and SN. As can be seen in Figure~5a, 
there exist variations of the order of 11 years. In fact, we get a reasonably
 good anticorrelation ($r = -0.27$, Student's $t = 3.28$, $p = 0.001$) 
between the slope (revised data) and SN, 
 suggesting that there  exists a relationship between the slope and 
activity during solar cycles. The  3--4 cycles
(30--40 years) periodic variations are relatively strong during the times of
 early solar cycles. 
 The existence of strong $\approx$11-year period variation in 
the cross-correlation coefficient (see Figure~5b) strongly indicates  
the existence of a strong relationship between the slope and solar cycle. 
The peak at $lag = 4$ implies that the slope leads  SN by  about four years. 
There is also a relatively large peak at $lag = 9$ (largest negative
 value of the cross-correlation coefficient) suggesting that the slope
 leads SN   by about nine years. The peak at $lag = 4$ suggests that 
 the slope around the preceding minimum of a solar cycle may be related to 
the strength of activity around the maximum of the same solar cycle.  
A large negative/positive (less negative)  slope at the 
minimum of a solar cycle  probably indicates that  the solar cycle 
 will have  a small/large maximum. This is some extent similar to 
the relationship between the strength of polar fields at minimum of a solar
 cycle and the solar cycle maximum (\opencite{sch78}). The peak at $lag = 9$ 
  suggests that a large negative/positive slope at the epoch  after 1--2-years
from the maximum of a solar cycle ($n$) indicates  a small/large 
amplitude for the next  solar cycle ($n+1$). This is some extent similar to the 
 relationship between the sum of the areas of the sunspot groups 
in the equatorial latitudes  just
 after about one-year from the maximum epoch of Solar Cycle~$n$ and 
the maximum of Solar Cycle~$n+1$ (\opencite{jj07}). 
We checked whether we can use the aforementioned  
  relationships of the slope and SN for predicting  the maximum 
of Solar Cycle~25. It is found impossible. That is, we get a very large  
uncertainty in the predicted value.
   The aforementioned relations of the slope and SN indicate that large
 poleward flows of magnetic flux (of decaying active regions) during 
the declining  phase of a solar cycle may enhance the strength of polar fields 
at the following minimum.  Large equatorward flows (including down flows at 
active regions) of 
magnetic flux may enhance the strength of activity in the equatorial latitudes  
just after one-year from the maximum of the solar cycle. 

\begin{figure}
\centering
\setcounter{figure}{0}
\begin{subfigure}
\includegraphics[width=7.0cm]{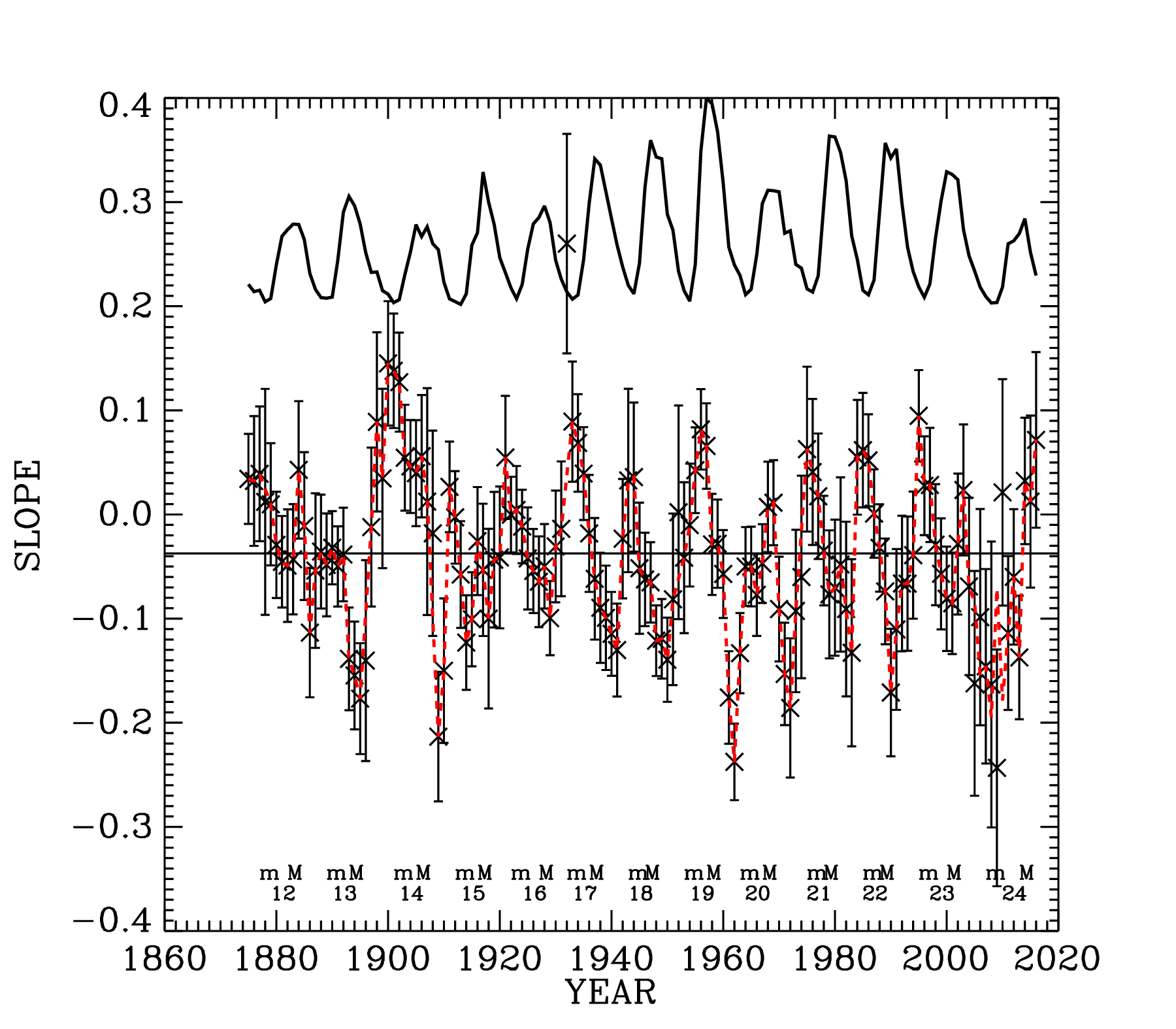}
\end{subfigure}
\begin{subfigure}
\includegraphics[width=5.0cm]{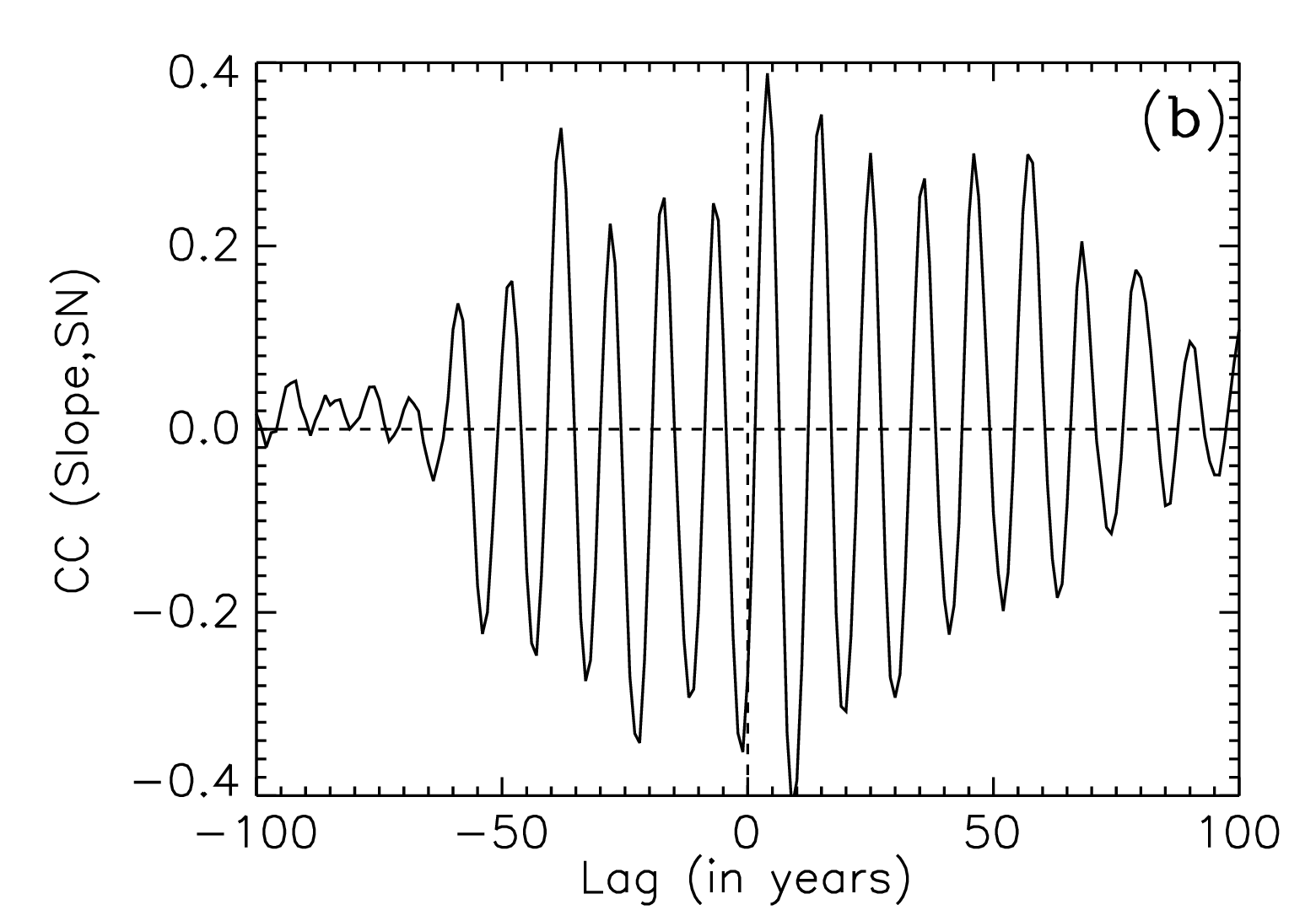}
\end{subfigure}
\setcounter{figure}{4}
\caption{({\bf a}) Plot of the values (cross) of the slopes of 
the linear best-fits of the meridional velocity 
($\langle v_{mer} (\theta) \rangle$)
 and residual rotation ($\Delta \langle v_{rot} (\theta) \rangle$)
determined from  sunspot group data in 3-year MTIs: 1874--1876, 
1875--1877,$\dots$,2015--2017, versus middle years of these intervals. 
The  dashed curve (red)
 represents the revised data after those values having
 $\sigma$ (standard deviation) greater than
 1.7$\times$the median $\sigma$ were replaced with mean of their respective
neighbor values. The continuous curve represents  the variation in
yearly mean values of SN.
 The Waldmeier numbers of the solar
cycles and the corresponding epochs of the minima and the maxima of solar 
cycles are also
 shown by the symbols $m$ and $M$, respectively.
The  horizontal continuous-line represents the mean value over the
whole period. ({\bf b}) Plot of the coefficient of cross-correlation (CC) 
between the slope and SN  versus lag.}
\label{f5}
\end{figure}

\begin{figure}
\centering
\includegraphics[width=11cm]{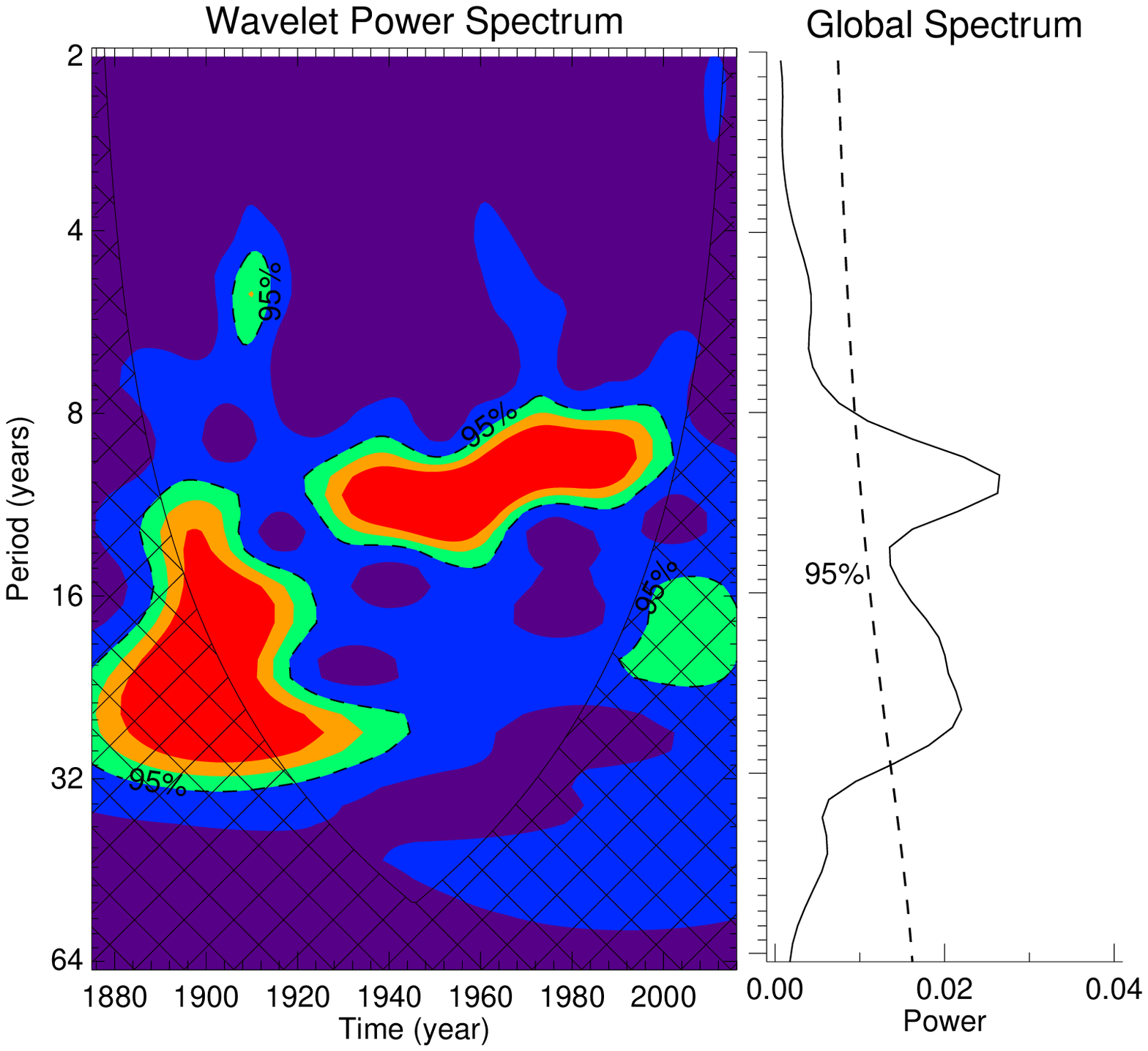}
\includegraphics[width=11cm]{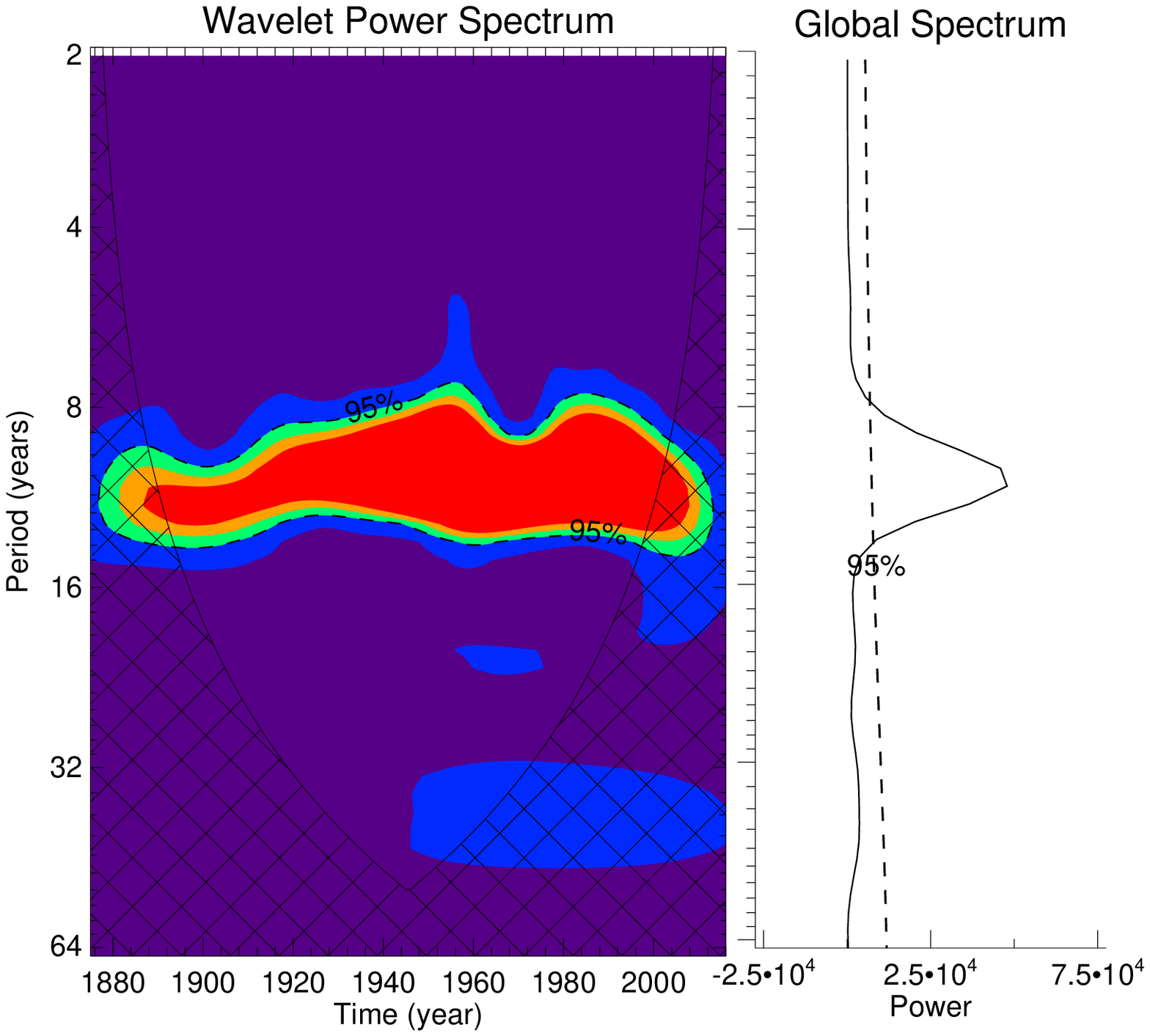}
\caption{(Upper panel) Wavelet power spectra
(left) and global spectra (right) of the slope of
linear relationship between the  meridional 
velocity ($\langle v_{mer} (\theta) \rangle$)
 and residual rotation ($\Delta \langle v_{rot} (\theta) \rangle$) of 
sunspot groups
 during 1875--2016  shown in  Figure~5a.
(Lower panel) Wavelet power spectra (left) and global spectra (right) 
of the annual mean  SN during 
1875--2016.
The wavelet spectra are  normalized
by the variances of the corresponding time series. The shadings are  at
the normalized variances of 1.0, 3.0, 4.5, and 6.0.
The dashed curves represent the 95\,\% confidence levels
deduced by assuming a white-noise process.
The cross-hatched regions indicate the cone of
influence where edge effects become significant (Torrence and Compo, 1998).}
\label{f6}
\end{figure}

\begin{figure}
\centering
\setcounter{figure}{0}
\begin{subfigure}
\includegraphics[width=5.5cm]{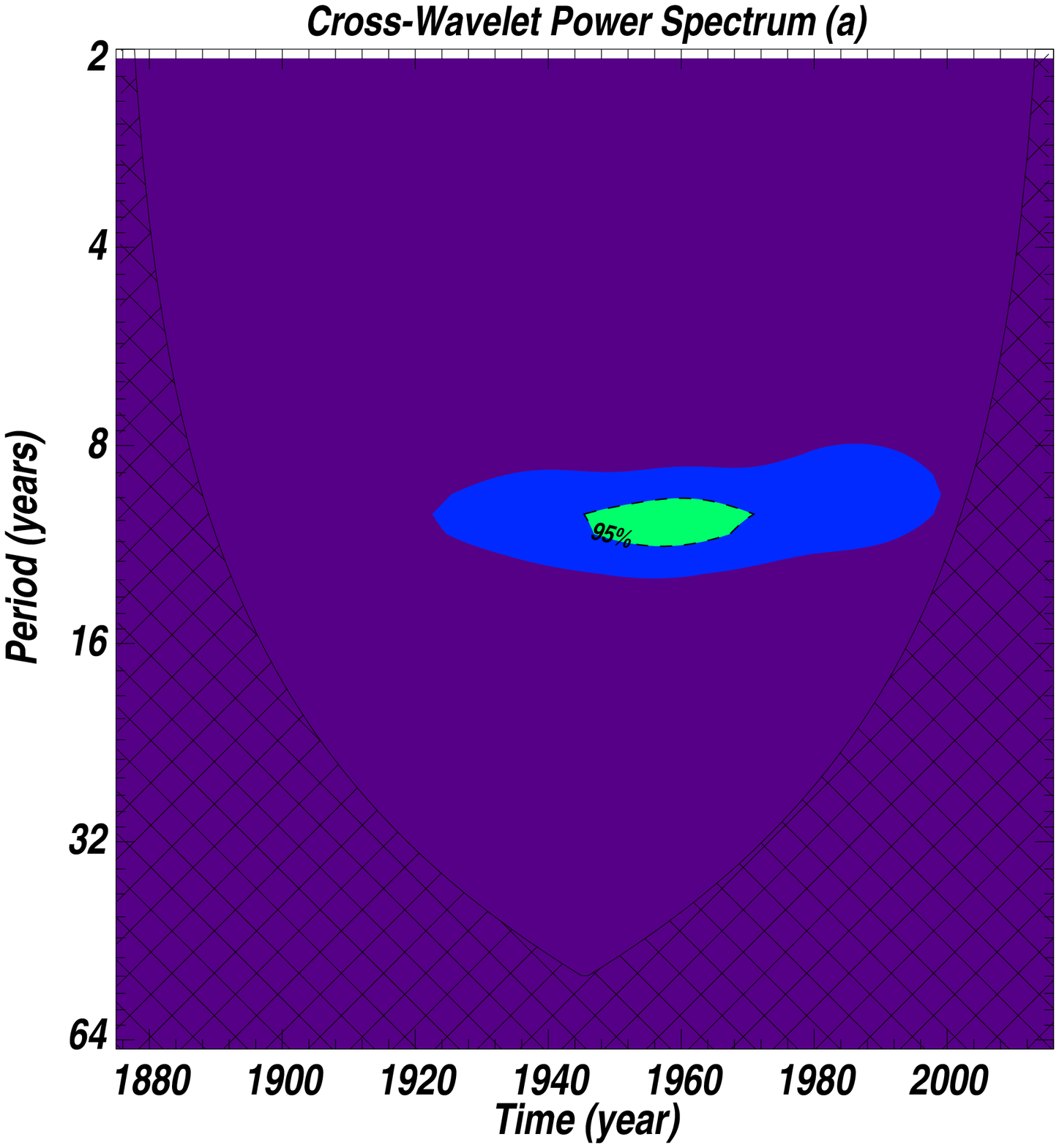}
\end{subfigure}
\begin{subfigure}
\includegraphics[width=5.5cm]{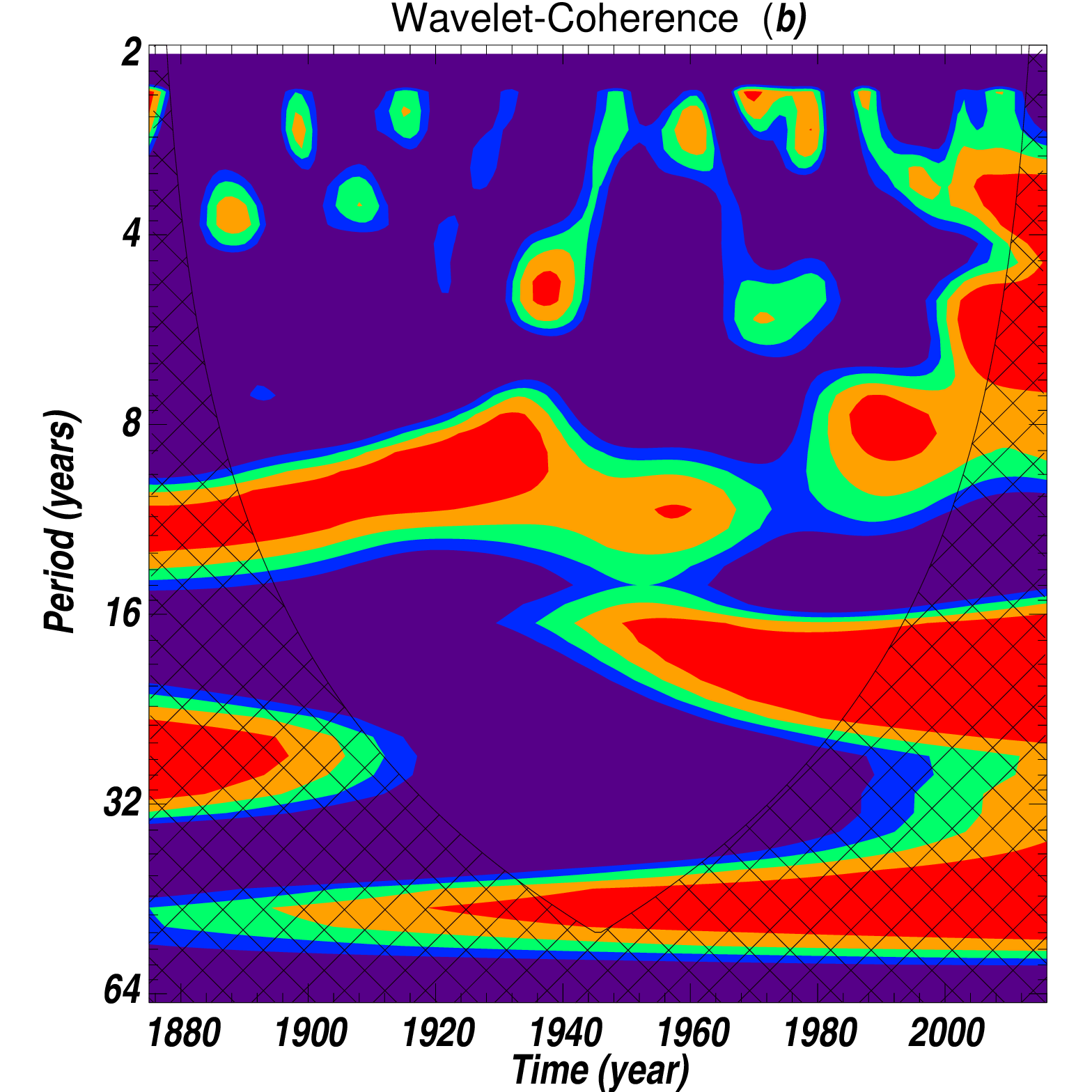}
\end{subfigure}
\setcounter{figure}{6}
\caption{Panels  ({\bf a})  and ({\bf b}) show the cross-wavelet
power spectrum  and wavelet-coherence spectrum, respectively,
 of sunspot number (SN)  and the slope of the linear relationship between 
 meridional velocity ($\langle v_{mer} (\theta) \rangle$) and 
residual rotation ($\Delta \langle v_{rot} (\theta) \rangle$)
 during the period 1875\,--\,2016. The shadings are at
levels 1.0, 3.0, 4.5, and 6.0.
The  cross-hatched regions indicate the cone of
influence where edge effects become significant (Torrence and Compo, 1998).}
\label{f7}
\end{figure}

\begin{SCfigure}
\centering
\vspace{-0.5cm}
\includegraphics[width=7.5cm]{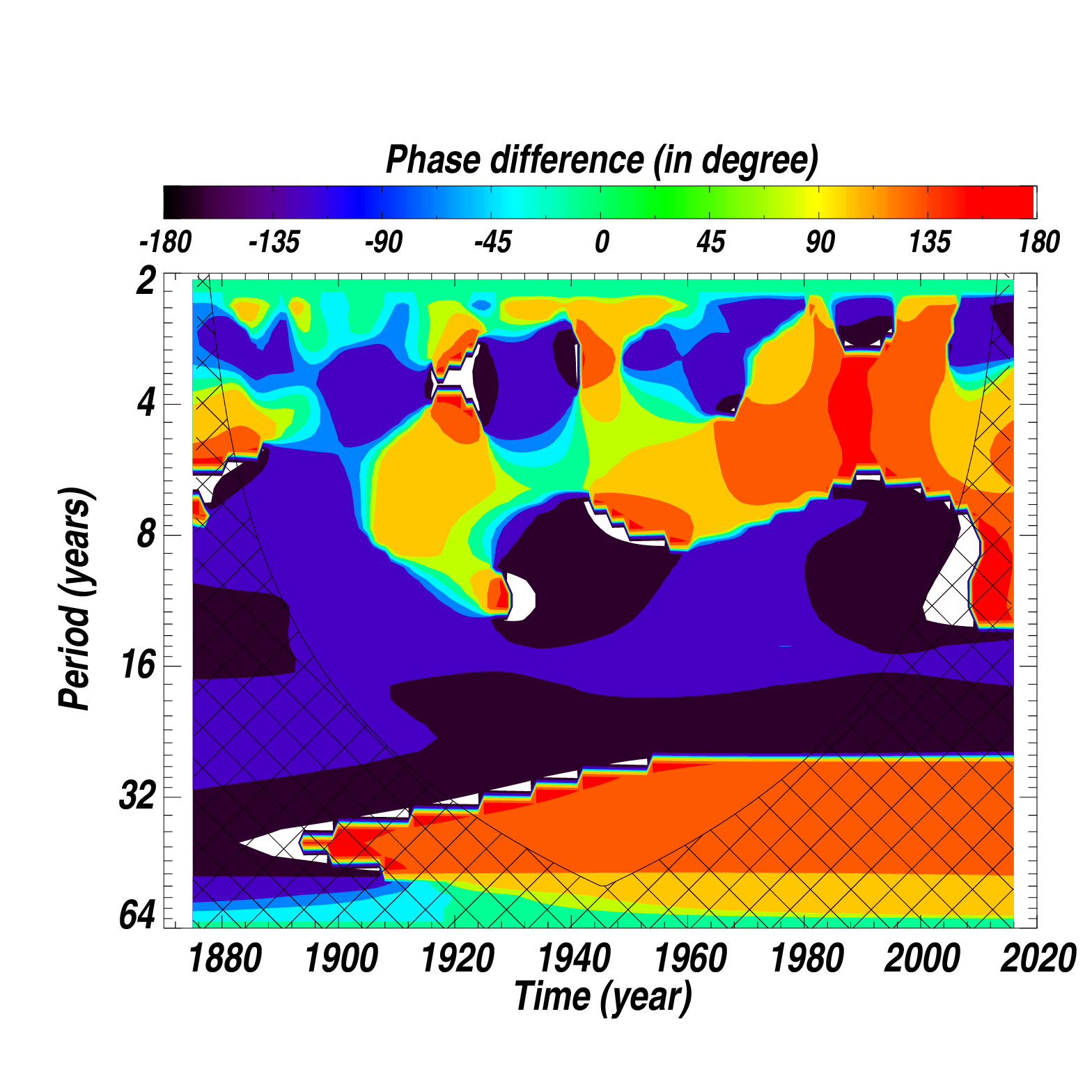}
\caption{Wavelet phase difference between  SN and the slope of the 
linear relationship between  meridional 
velocity ($\langle v_{mer} (\theta) \rangle$) and residual rotation 
($\Delta \langle v_{rot} (\theta) \rangle$) in 3-year MTIs during the 
period 1875\,--\,2016.
The  cross-hatched regions indicate the cone of
influence where edge effects become significant (Torrence and Compo, 1998).}
\label{8}
\end{SCfigure}

 In Figure~6 we compare the Morlet wavelet spectra of the 
variations in  the slope and SN shown in Figure~5a. Obviously, 
11-year periodicity 
is strongly present in SN throughout the analyzed data window.  
 This periodicity seems to be present in the slope almost throughout the
data window, but it was very weak during 1920--1940. 
The 16--32-year periodicities were strongly 
present in the slope during the time of early solar cycles. Figure~7 shows the 
cross-wavelet and wavelet-coherence spectra of the slope and SN.
 The cross-wavelet spectrum (Figure~7a)  suggests that  there exists  
a statistical significant similarity in the temporal behaviors of the slope 
and SN only during the period 1950--1970, because outside this interval the 
slope  has slightly smaller than 11-year periodicity (see Figure~6a). 
  In Figure~7b    
there is a suggestion on the existence of strong coherence 
between the $\approx$11-year variations of the slope and SN 
 before 1940 and 1980-onward. There is also a suggestion on the 
existence of coherence between the  20--30-year periodicities of 
 the slope and SN from 1940-onward. 
There are  episodes  of  3--5-year periodicities 
in both the slope and SN  between 1880 and 2016.
There is a suggestion of   
the coherence in  40--50-year periodic variations  in  the slope and SN, 
but this signal  is within  the area of cone-of-influence. That is, 
the signal of this periodicity is not well resolved, hence this periodicity 
 cannot be detected here due to inadequate data.  Figure~8 shows the 
 spectra of the wavelet-phase difference of the slope and SN. As can be seen in
 this figure, there exists $\approx -180^\circ$ phase difference between 
the 11-year period variations of the slope and SN before 1900 (the slope 
seems to be lead SN), during the periods 1940--1960 and  1980--2000.
 A 20--30 year period variations   
of the slope and SN seems to be having   $\approx$ $ -180^\circ$ phase
 difference 
throughout 1875--2016. This is consistent with the anticorrelation between 
the slope and SN found above.  There seem to be    considerable phase 
differences 
 also in the aforementioned remaining all periodic variations. A $90^\circ$ 
difference indicates that there is a considerable phase mixing in 
the corresponding variations of the slope and SN.

\subsection{Cycle-to-Cycle Variation in the Slope}
In Table~2 we have given the  values of the intercept  and the 
slope of the linear
relationship  between  $\langle v_{mer} (\theta)\rangle$  and  
$\Delta\langle v_{rot} (\theta) \rangle$
 determined from the data during each of  Solar Cycles 12--24, and also 
 from the combined data of all cycles. The corresponding
values of the correlation coefficient, Student's t, 
and probability  are given. The  cycle interval, number of
data points ($N)$,  and maximum version-2 sunspot number ($R_{\rm M}$) 
 are also given. In Figure~9 we have shown the  
cycle-to-cycle variations in
the  slope  during Solar Cycles 12--24. As can be seen in this figure,  
 although in several solar cycles the correlation is poor and the 
coefficients of linear best-fits have large  uncertainties (also see Table~2),
there is a possibility of a considerable variation in the slope on the time 
scale of about 3--4-cycles, suggesting the existence of a 
3--4-cycle periodicity in the slope. 
 In Solar Cycles~15, 18, 19, 20, and 22 the slope has
 significant negative values. The value of  Solar Cycle~24 is 
also some extent negative (this cycle is incomplete). 
 In Solar Cycles  14, 17, and 21 
the slope has positive values, but these values are not significantly 
different from zero. 
 The equatorward  angular momentum transport may be relatively 
much less (absent) in Solar Cycles
 14, 17, and 21. The correlation  ($r = -0.28$) between  the 
slope and the amplitude of solar 
cycle is found to be insignificant, indicating that there is no relationship 
between the slope and strength of activity on a long-time scale (longer than 
 11-year period).

\begin{table}
{\tiny
\caption[]{The values of intercept ($C$)  and slope ($D$) of the linear 
relationship between $\langle v_{mer} (\theta) \rangle$ and  
$\Delta\langle v_{rot} (\theta) \rangle$ determined from 
the data during the whole period of each  solar cycle and 
 from the combined data of all Solar Cycles 12--24 (the last row).
 The corresponding     
values of the correlation coefficient ($r$), Students' t ($\tau$) 
and probability ($Prob.$) are given.  Cycle interval (Time), number of 
data points ($N$),  maximum ($R_{\rm M}$) Version-2 sunspot numbers are 
also given.}
\begin{tabular}{lcccccccc}
\hline
Cycle&Time& $R_{\rm M}$& $C$ & $D$& $ r$ & $\tau$& $Prob.$&  $N$\\
12&  1878--1889& 124.4&$  0.781\pm   0.819$&$  -0.023\pm   0.027$&$ -0.09$&   0.85&  0.802& 101\\
13&  1890--1901& 146.5&$ -0.398\pm   0.859$&$  -0.028\pm   0.028$&$ -0.10$&   0.97&  0.834& 107\\
14&  1902--1912& 107.1&$  0.075\pm   0.871$&$   0.029\pm   0.029$&$  0.11$&   1.01&  0.843&  87\\
15&  1913--1922& 175.7&$ -1.422\pm   0.722$&$  -0.086\pm   0.030$&$ -0.29$&   2.85&  0.997&  93\\
16&  1923--1932& 130.2&$ -0.381\pm   0.724$&$  -0.026\pm   0.024$&$ -0.11$&   1.05&  0.853& 103\\
17&  1933--1943& 198.6&$ -0.388\pm   0.689$&$   0.006\pm   0.026$&$  0.02$&   0.22&  0.587& 111\\
18&  1944--1953& 218.7&$ -1.000\pm   0.718$&$  -0.086\pm   0.029$&$ -0.27$&   2.97&  0.998& 113\\
19&  1954--1963& 285.0&$  0.074\pm   0.698$&$  -0.047\pm   0.024$&$ -0.18$&   1.99&  0.976& 119\\
20&  1964--1975& 156.6&$  1.064\pm   0.648$&$  -0.048\pm   0.023$&$ -0.18$&   2.07&  0.980& 127\\
21&  1976--1985& 232.9&$ -0.436\pm   0.787$&$   0.018\pm   0.031$&$  0.05$&   0.56&  0.712& 119\\
22&  1986--1995& 212.5&$  0.909\pm   0.724$&$  -0.050\pm   0.030$&$ -0.16$&   1.69&  0.953& 113\\
23&  1996--2007& 180.3&$  1.383\pm   0.664$&$  -0.017\pm   0.030$&$ -0.05$&   0.56&  0.713& 133\\
24&  2008--2017& 116.4&$  0.476\pm   0.949$&$  -0.053\pm   0.043$&$ -0.13$&   1.23&  0.889&  88\\
All &1878--2017&    &$    0.16\pm    0.20$&$   -0.028\pm   0.008$&$ -0.1$&    3.73&  0.9999&1388\\
\hline
\end{tabular}
\label{table2}
}
\end{table}

\begin{SCfigure}
\centering
\includegraphics[width=7.5cm]{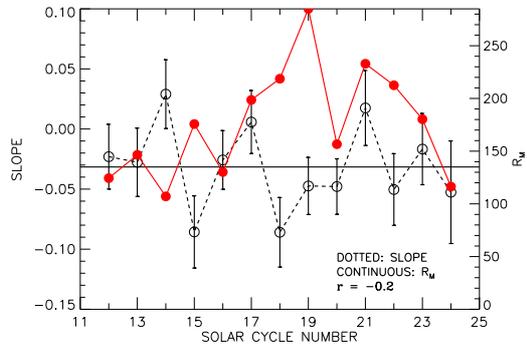}
\caption{Cycle-to-cycle variation in 
the  slope (open circle-dotted curve) of the linear best-fit of 
meridional velocity ($\langle v_{mer} (\theta) \rangle$) and 
$\Delta\langle v_{rot} (\theta) \rangle$ 
 of sunspot groups during Solar Cycles 12--24. 
The filled circle-continuous curve (red) represents the cycle-to-cycle  
variation  in solar-cycle  amplitude ($R_{\rm M}$).}
\label{f9}
\end{SCfigure}

\section{Conclusions and Discussion}
We analyzed  the combined  142 years GPR and DPD sunspot-group data 
 and found that there is a considerable latitude--time dependence in 
 the yearly mean residual rotation rate of sunspot groups.  
The yearly average residual rotation rate  is roughly from
 $-120$ m $s^{-1}$ to $80$ m $s^{-1}$.
 In a large number of solar cycles the
rotation is some extent weaker during maxima than that of during minima.
In many solar cycles there is an  indication that bands of residual rotation
rate migrate from high  to low latitudes  over 8--10-year periods.
The analysis of the data during 
Solar Cycles 12--24 that are folded  according to the years from their
respective epochs of maxima suggests
 the existence of  alternate bands of  slower  and faster than 
average rotation    within the activity
belt.   A  $\approx 10^\circ$-wide 
 slow band ($-$70 to $-80$ ms$^{-1}$) 
 seems to be
 originated around $35^\circ$ latitude and a  narrow one (only 
$5^\circ$ wide)  seems to be
 originated around $15^\circ$ latitude during the minimum of
 a solar cycle and both migrated toward low latitudes.
 However, their 
relationship with the well-known torsional oscillations is not clear.
The maximum absolute value of the residual rotation rate 
 is much larger than the amplitude of the
 velocity as well as the  magnetic  torsional oscillations.

There is also a
considerable latitude--time dependence in the yearly mean meridional 
motion of sunspot groups. There
are alternate bands of equatorward and poleward
motions. Overall, the equatorward motion is dominant with
velocity 8--12 m s$^{-1}$ mostly around maxima of more solar cycles, whereas
the poleward motion seems to be relatively weak
with velocity  only 4--6 m s$^{-1}$  mostly around minima of
more solar cycles. 
 The analysis of the folded  data of
 Solar Cycles 12--24 suggests  no clear  
equatorward or poleward migrating 
   bands of meridional motions.

 There exists a statistically
significant anticorrelation between the meridional motion and residual 
rotation. The  corresponding linear-least-squares best-fit is
 reasonably good in the sense
that the slope  ($ -0.028 \pm 0.008$) is about 3.5 times larger 
than its standard
deviation. As per the sign convention used for the meridional velocities of 
sunspot groups, the significant  negative value of the slope
 can be considered as a measure of
the strength of angular momentum transport toward equator.
 This result is in a qualitative agreement with previous results for 
sunspots (\opencite{sudar14}; \opencite{sudar17}). However, our result 
has a lower amplitude, which might be a consequence of different selection
process. We can conclude that the results of present work generally  confirm 
earlier results (\opencite{sudar14}; \opencite{sudar17}) and represent, 
together with them, a strong evidence that Reynolds stress is indeed  the 
dominant mechanism for maintaining solar differential rotation via angular 
momentum transport towards equator.

 The cross-correlation between the  slopes determined from the data
in 3-year moving time intervals and yearly mean
 sunspot number (SN) suggests that the slope leads SN by about 4 and 9 years.
 The Morlet wavelet spectrum of  SN suggests, obviously, 11-year periodicity
is strongly present in SN throughout the analyzed data window.
 The Morlet wavelet spectrum of the slope also   suggests  the 
existence  of $\approx$~11-year periodicity in the slope 
 almost throughout the data window, but it was very weak during 1920--1940.
The cross-wavelet spectrum  suggests that  there exists
a statistical significant similarity in the temporal behaviors of the slope
and SN only during the period 1950--1970. The wavelet-coherence  spectrum 
 suggests that  there  exists a  strong coherence between 
the $\approx$~11-year variations of the slope and SN before 1940 and
1980-onward.   There exists $\approx$~$-180^\circ$ phase difference between
the 11-year period variations of the slope and SN before 1900,
during 1940--1960, and during 1980--2000. 
The  20--30-year period variations
of the slope and SN seem to be having   $\approx$~$-180^\circ$ phase difference
throughout 1874--2017. However, this periodicity is very weak in both the 
slope and SN.  The  above mentioned result that the slope leads SN by
 about 4 and 9 years is consistent with 
the $180^\circ$ phase difference between the $\approx$~11-year 
variations  and that between the $\approx$~20-year variations  of the
 slope and SN. Overall these
results suggest there exits a strong relationship between the slope
and amount of activity during a solar cycle. However,
the correlation  between  the cycle-to-cycle
modulations in the slope and the amplitude of solar
cycle is found to be insignificant, indicating that there is no relationship
between the slope and strength of activity on a long-time scale (longer than
 11-year period).

Reynolds stress is produced by interaction of convective elements and the 
Coriolis force which cause a correlation between longitudinal and 
latitudinal velocity components. 
The motion of sunspots represents the motion of somewhat deeper layer of
the sun (see \opencite{jj13}). Recent local helioseismic
 studies suggest 
that patterns of variations in the large scale flows are the same in both 
quite and active  regions and the  flow patterns in active regions 
associated to a deeper layers than those of quite regions 
(\opencite{komm20}).  The anticorrelation between 
the meridional motion and residual rotation 
of sunspot groups found here implies the existence of couplings between 
these motions, 
somewhere in the convection zone. Since Reynolds stress produced can
 maintain the equatorial angular momentum transport and 
thus differential rotation (\opencite{gilm86}), hence  there exists
 correlation between the strength of activity
 and the slope of linear relationship between the meridional motion and
 residual 
rotation during a solar cycle. The existence of a $\approx 180^\circ$  
phase difference between 
the slope and activity during their 11-year and 20--30-year
 periodic variations implies the 
existence of couplings between the strength of magnetic activity and the
 angular momentum transport toward equator on these time scales.  
 However, our analysis suggest that there  is
no such relationship  on  longer than these time scales.

\section{Acknowledgments}
 The author thanks the anonymous reviewer for useful
 comments and suggestions. 
The author acknowledges the work of all the
 people contribute  and maintain the GPR and DPD  Sunspot databases.
The sunspot-number data are provided by WDC-SILSO, Royal Observatory of
Belgium, Brussels. The wavelet software was provided by C. Torrence and
 G. Compo and is available at {\sf paos.colorado.edu/research/wavelets}.

\section{Conflict of interest}
The authors declare that they have no conflicts of interest.

{}
\end{article}
\end{document}